\documentstyle [11pt,epsf] {article} 
\begin{document}
\begin{titlepage}
\begin{center}
\vspace{2cm}
\Large 
{\bf Understanding the Astrophysics of Galaxy Evolution: the role of
spectroscopic surveys in the next decade}
\\                                                     
\vspace{1cm} 
\large
Eric Bell, Marc Davis, Arjun Dey, Pieter van Dokkum, Richard Ellis,  Daniel Eisenstein,
 Martin Elvis,  Sandra Faber, Carlos Frenk, Reinhard Genzel, 
Jenny Greene, Jim Gunn,  Guinevere Kauffmann, Jill Knapp,  Mariska Kriek, 
James Larkin, Claudia Maraston, 
Kirpal Nandra, Jerry Ostriker,  Francisco Prada, 
 David Schlegel, Michael Strauss,  
Alex Szalay, Christy Tremonti, Martin White, Simon White, Rosie Wyse \\
\vspace{0.5cm}
\normalsize
\begin {abstract}
Over the last decade optical spectroscopic surveys have characterized the
low redshift galaxy population
and uncovered populations of star-forming galaxies back to $z\sim 7$. This
work has shown that the primary epoch of galaxy building and black hole growth 
occurs at  redshifts of 2 to 3. The establishment in
this same decade of the concordance $\Lambda$CDM cosmology  shifted the
focus of galaxy population studies from constraining cosmological parameters
to characterizing  the  processes which regulate the formation and
evolution of galaxies. 
In the next decade, high redshift observers will attempt to 
formulate a coherent evolutionary picture
connecting galaxies in the high redshift Universe to galaxies today. 
In order to {\em link galaxy populations at different redshifts}, we must not
only  {\em characterize their  evolution in a systematic way},  
we  must {\em establish which physical processes
are responsible} for it. Considerable progress has already been
made in understanding how galaxies evolved from  $z \sim 1$ to the present day.
Large spectroscopic
surveys in the near infrared are required to push these studies back towards 
the main
epoch of galaxy building. Only then will we  understand 
the full story of the formation of $L_*$ galaxies like our own Milky Way.
A large near-IR spectroscopic survey will also provide the
calibration needed to avoid systematics in the large photometric programs
proposed to study the nature of dark matter and dark energy. We provide an
outline design for a multi-object 0.4 to 1.8 micron spectrograph, which could
be placed on an existing telescope, and which would allow a full 
characterization of
the galaxy population out to $z\sim 2$.
We strongly recommend a serious further study to                
design a real instrument, which will be required for  galaxy
formation studies to advance to the next frontier.
\end {abstract}
\vspace{0.8cm}
\end{center}
\normalsize
\end {titlepage}

\section {Current state of observations of high-redshift galaxies}

Over the past decade and a half, our observational understanding of galaxy
evolution has grown enormously.  Steidel and collaborators demonstrated that
colour selection techniques allow so-called Lyman Break galaxies (LBGs) to be
isolated efficiently at redshifts $\sim$ 3, thus breaking the $z=1$ redshift
barrier that had dogged the field for many years.  Spectroscopy then confirmed
the redshifts and showed that LBGs had moderate mass ($10^{9}-10^{10}
M_{\odot}$) and metallicity (0.3 solar), that they were forming stars very
rapidly, and that supernovae were driving significant outflows.  Since then,
the race to claim the record for the highest redshift object has absorbed many
in the field. Today, this record stands at $z=6.96$. These distant objects
appear as tiny smudges in the deepest HST imaging data obtained and have low
masses. At $z\sim 5$ the integrated mass density in stars was less than 10
percent of its present day value.  Although galaxies at $z>6$ may hold the key
to understanding the re-ionization of the Universe, their story is not central
to our understanding of the formation of the main components of typical spirals and ellipticals in
the nearby Universe.

Today, we know that typical $L_*$ galaxies assemble at $z = 2$ to 3. These are
the redshifts where star formation and black hole accretion activity peaked.
Lyman Break galaxies were initially viewed as the obvious progenitors of
present-day $L_*$ galaxies.  However, it is now clear that although the Lyman
break technique is extremely powerful, it does not pick up {\em all} high
redshift galaxies.  Franx and his collaborators found a significant population
of very red galaxies (DRGs) that did not satisfy the LBG colour
selection. Some are red because they were dusty, but a significant fraction
are red because star formation had already switched off some time
previously. DRGs are more massive than LBGs ($> 10^{10} M_{\odot}$) and denser
than elliptical galaxies in the local Universe.

The Great Observatories, HST,Chandra and Spitzer, also spent considerable
effort on deep surveys, as did ground-based observatories such as the JCMT.
Each new survey successfully found galaxies at $z \sim 2$ and triumphantly
announced the discovery of a "new population" with its own acronym. These now
fill the literature on high redshift galaxies with a bewildering menagerie:
XBONGs (X-ray bright optically normal galaxies) from Chandra, DOGs
(dust-obscured galaxies) from Spitzer, 
UVLGs (ultra-violet luminous galaxies) from GALEX, not to mention LBGs, DRGs,
BXs, BzKs and BMs from ground-based optical and near-IR surveys.

This situation is symptomatic of a subject that is still in its infancy.  Over
the next decade, high redshift observers will attempt to unify these different
galaxy classes, in order to formulate a coherent evolutionary picture
connecting galaxies in the high redshift Universe to galaxies today.  Just as
evolutionary biology progressed from Linnaean taxonomy to a more mature phase
where the theories introduced by Charles Darwin were used to connect and to
unify the different species, so too must the field of galaxy evolution
progress beyond the naming of the animals.  As we will describe, large
spectroscopic surveys of galaxies in the near infrared are critical if we are
to make this transition successfully.

\section {Current state of the theory of galaxy formation}

The scientific development of the last decade which most impacted galaxy
formation theory was the establishment of the $\Lambda$CDM concordance model
of cosmology, in which the Universe consists of 70\% dark energy, 25 \% dark
matter and 5 \% ordinary matter.  Before this happened, it was thought that
galaxies might provide strong constraints on cosmological parameters.
Considerable attention was devoted to the abundances, masses and ages of
high-redshift galaxies, because these properties constrain the epoch of
structure formation, and hence parameters such as $\Omega$ and $H_o$.  The
precise cosmological parameters provided by the cosmic microwave background,
supernovae and low redshift large-scale structure 
data have now made such approaches obsolete.

For the galaxy formation theorist, the era of precision cosmology means that
the challenge now shifts to understanding the formation of the galaxies
themselves.  The evolution of the main matter component of the Universe, the
dark matter, can be modelled to high accuracy by running N-body simulations on
large supercomputers. The same does not hold for the baryons.  It has become
increasingly clear that the processes that regulate cooling, condensation and
star-formation at the centers of dark matter halos are extremely complex. It
is likely that we will never be able to model them ``ab initio" using direct
simulations in the way we now model dark matter evolution. One of the main
problems hampering progress is the enormous dynamic range in scale that must
considered when treating star and galaxy formation.

Some of these processes, such as the effect of supernova explosions on the
interstellar medium of the galaxy, have been discussed for many years.  Others
are just now coming to be appreciated.  The discovery that the centers of
essentially all luminous nearby galaxies harbor a supermassive black hole with
mass tightly related to that of the surrounding bulge, has revolutionised
current thinking about how giant elliptical galaxies form. Rather than being
simply an end-product of galaxy formation, supermassive black holes are
thought by many to play an active role in regulating the formation process. 
In the
most extreme form, ``feedback" from an accreting black hole is supposed to
blow all the gas out of the galaxy and its surrounding dark matter halo, thus
shutting down star formation and terminating growth.  Except in the cores of
rich galaxy clusters, the observational evidence for such violent AGN feedback
remains sparse, and the models suggesting its feasibility are necessarily
schematic. Nevertheless, AGN feedback is currently one of the hottest topics
in galaxy formation theory.

The main reason for this is that the $\Lambda$CDM cosmological model provides
robust predictions for the dark matter halos of typical giant ellipticals and
for the mass of baryons they should contain. Simple physics predicts that in
the absence of compensating effects, the gas should cool and form many more
stars than are observed.  {\em It is this tension between theoretical
  predictions and what is observed in the real Universe, that will continue to
  drive both observational and theoretical progress in the field of galaxy
  evolution for the foreseeable future.} Considerable theoretical effort will
be needed to develop more accurate sub-grid models to describe star formation,
chemical enrichment, and supernova and AGN feedback. Such sub-grid models
must be included in larger scale simulations of the evolution of galaxy
populations so that comparison with observations can determine the extent to
which the effort has been successful.

\section{From taxonomy to  evolutionary science: 
the role of large galaxy surveys}

The main goal in galaxy formation for the next decade will be to {\em connect}
galaxy populations observed at different redshifts within the framework of a
single, consistent picture of galaxy evolution.  We will now argue that large
near-infrared spectroscopic galaxy surveys are crucial if this effort is to
succeed.

In order to understand galaxy evolution, it is critical to select galaxies
using photometric passbands that sample the light emitted by the population of
long-lived stars, i.e. longwards of the rest-frame 4000 \AA\ break, but
shortwards of wavelengths where emission by thermally-pulsating giant (TP-AGB)
stars or by dust becomes important.  In contrast to the light emitted at
ultra-violet wavelengths or far-infrared wavelengths, which is dominated by
(direct or reprocessed ) radiation from short-lived O and B stars, or
radiation at X-ray or radio wavelengths, which is influenced by processes
occurring in the immediate vicinity of the black hole, the rest-frame optical
light from a galaxy is normally stable over relatively long timescales.  A
survey selected in the {\it rest-frame} optical will allow us to characterize
different transient phenomena (e.g. starbursts and AGN) occurring in the
galaxy population at a given epoch.  It will also allow us to connect galaxy
populations observed at different epochs and to estimate galaxy stellar
masses, the quantity that is likely most tightly correlated with dark matter
halo mass and so the best link to the underlying cosmological model.

A survey must be large ($\sim$ few $\times 10^5$ galaxies) in order to
disentangle {\em covariances} in the physical properties of galaxies. One
reason it is so difficult to understand how galaxies form is because almost
all galaxy properties are correlated.  Galaxy mass correlates with
morphological type, with colour, with metallicity, with star formation rate,
with gas content and with local and large-scale environment. The correlation
of property A with property B does not establish that B regulates A.  With a
large survey, one can control many properties at the same time , i.e. one can
look at how property A depends on property B if properties C through F are all
held fixed. 
Accurate measurement of the of the correlation function out to the  scales
needed to constrain dark matter halo masses typically requires samples of $10^4$ galaxies or more.
These techniques became standard analysis procedure with the advent of the
Sloan Digital Sky Survey; imaging and spectroscopy of well over half a million
galaxies at $z \sim 0.1$ have given us a detailed view of the astrophysical
processes at work in the low-redshift Universe. The next step is to establish
the evolutionary sequence that produced today's population of galaxies. Deep
optical surveys have made substantial progress in studying evolution at
$z<1$ but new efforts at longer wavelengths are needed to cover the
principal epoch of galaxy growth which lies at $z>1$. 

Two steps must be accomplished if one is to use a large survey to connect
galaxy populations at different redshifts: 1) The evolution of galaxies must
be {\em characterized} in a systematic way.  This can be done to some extent
using photometric surveys.  2) We must then establish which physical processes
are responsible for the observed evolution.  This requires large spectroscopic
surveys in the near infrared.

\section {Imaging and  spectroscopy: what can be accomplished}

\subsection {Imaging only}

One cannot begin to study the evolution of galaxies unless one has some idea
of the redshift at which they lie.  With broad-band imaging data alone, one
must resort to photometric redshifts. The standard method is to fit the
photometry to a set of template spectra, drawn from population synthesis
models or using the observed SEDs of low-redshift galaxies. The problem lies
in the appropriate choice of templates. The galaxy population evolves
strongly, so templates constructed from low-redshift galaxies do not
necessarily match very well at high redshift.  If one uses models, then one
has to worry whether these are a good representation of real galaxies, how to
introduce corrections for dust, and so on. The best results are obtained if
one has a large training set of redshifts derived from real galaxy spectra.
The typical errors on photometric redshifts using spectroscopic training sets
and state-of-the-art techniques such as neural nets, are $\sigma
(\Delta(z)/(1+z))=0.06$ at $z>1.5$  for redder galaxies with strong 4000 \AA\ breaks. The errors
are larger for blue galaxies, which have featureless spectra dominated by
emission lines that vary strongly from object to object.

Despite these limitations, the current generation of large imaging surveys
(e.g.  COSMOS) has been able to characterize how the luminosity functions of
different galaxy populations (e.g. red/blue sequence galaxies, morphologically
early/late type galaxies, radio-loud and X-ray detected AGN) evolve out to
redshifts $\sim 1$.  The photo-z errors are too large to allow identification
of large-scale structure in redshift space, but with high quality imaging,
characterization of the underlying dark matter density field becomes feasible
through weak gravitational lensing.  By stacking many galaxies with similar
properties and measuring the tangential distortion of background galaxies, one
can directly measure the mean mass profile of the galaxies' dark matter halos.
This provides extremely powerful constraints on galaxy formation models, and
the next generation of large imaging surveys (Pan-Starrs, DES, LSST) will
bring real advances in understanding here. If we are to realize the full power
of the lensing techniques, however, we need to stack the galaxies according to
a variety of different {\em physical} properties, and to calibrate accurately
the photo-z's of the background sources. This is where near-infrared
spectroscopy will be crucial.

\subsection {Imaging and optical spectroscopy}

Galaxy redshifts out to $z \sim 1.4$ can be obtained from optical spectra. At
higher redshifts, the doublet [OII]$\lambda 3727$ is no longer accessible with
standard optical spectrographs and one enters the so-called ``redshift
desert".

Over the past 5 years, surveys such as DEEP2 and VVDS have amassed a few tens
of thousands of optical spectra of galaxies at $z \sim 1$.  We have learned
that the standard Hubble sequence is still largely in place at these
redshifts, but that there has been significant evolution in both late- and
early-type galaxy populations.  At $z\sim 1$ the total stellar mass in red
sequence galaxies is only about a third of the present-day value, and the mean
star formation rates of blue sequence galaxies are larger by a factor of 7 to
10 in the mean.  Spectral information has been critical in attempts to
elucidate what is {\em causing} this evolution.

Accurate redshifts have allowed the construction of group and cluster catalogues
and reliable estimates of local galaxy density. From this, we have learned that the
relation between galaxy color and environment evolves very strongly out to
$z=1$. Locally there are no massive star-forming galaxies in very rich
environments, but at $z=1$ these are apparently quite common.

The kinematics of galaxies at $z=1$ were also  studied
using the higher resolution spectra provided by DEEP2. From this  we learned
that galaxies that look like ordinary spiral galaxies in the HST
images
frequently have  velocity fields that are dominated by
dispersion ($V/\sigma < 1$) rather than rotation. These are currently
hypothesized to be galaxies where the infall of cold, clumpy gas  
has produced turbulent, disordered velocity fields. 

Emission lines in the optical spectra of these
intermediate redshift   star-forming galaxies 
provided valuable information on the physical conditions in their interstellar
media.  The evolution of the mass-metallicity relation with redshift was
studied in detail, the main result being that the mean gas-phase metallicity
of star-forming galaxies appears to drop quite strongly with redshift at fixed
stellar mass, once again supporting the idea that the blue galaxy population
at high redshifts might be experiencing a rather high rate of gas accretion.

Rather little is currently  known about the stellar metallicities and element
abundance ratios of high redshift
galaxies. Stellar absorption lines do contain important information, but
individual high-redshift spectra generally have too low S/N to permit
metallicity determinations. However, one can get quite far by stacking the
spectra of many similar galaxies.

Studies of stellar absorption lines in high redshift galaxies are just
beginning. One very exciting development has been the identification of a
population of extreme post-starburst galaxies at redshifts $z \sim 1$. These
galaxies have extremely strong Balmer absorption lines, but weak or absent
emission lines, indicating that they must have experience a very significant
burst of star formation in the past gigayear. One pressing question is whether
these galaxies are in some way connected with the population of galaxies with
actively accreting black holes at the same redshift. Progress on that question
has suffered because the emission lines needed to diagnose the presence of an
AGN (H$\alpha$ and [NII]$\lambda$6548) are already redshifted into the near-IR
part of the spectrum at $z\sim 1$, and hence are inaccessible with current
instruments.

\subsection {Why we need near-IR spectroscopic surveys in the next decade}

The reason is very simple.  As we push out to redshifts $z\sim2$ when
typical spiral galaxies and ellipticals were in their most active
formation phase, optical spectra sample the rest-frame ultraviolet
part of the galaxy SED which is dominated by light from the most
massive, short-lived stars.   There are very few strong stellar
absorption lines at these wavelengths, particularly in the near-UV,
and no important nebular emission lines apart from Ly$\alpha$, which
is difficult to interpret due to complex radiative transfer effects.
It is therefore difficult to measure even the most basic galaxy
properties -- redshifts and velocity dispersions.   The UV does
provide an important shapshot of a galaxy's recent star formation
activity, but to realize its full diagnosic power data at longer
wavelengths are required to constrain the effects of dust attenuation.
  The UV also contains a large number of interstellar absorption lines
that can provide important diagnostics of the kinematics and
ionization state of the ISM, but optical data is needed to relate the
properties of these absobers to the star formation, AGN activity, and
metallicity of their hosts.

If we are to understand the full story of the formation and evolution of our
home galaxy, the Milky Way, and its peers, we require large near-IR
spectroscopic surveys. 

\section {Cosmology and spectroscopy}

In the next decade observational cosmologists will attempt to constrain the
nature of dark matter, dark energy and the (perhaps inflationary) processes
which generated structure. Large-scale spectroscopic surveys on large
telescopes will be critical to achieving reliable results in all these areas.

The distribution of dark matter is closely related to its nature, and can be
constrained through dynamical studies of clusters and satellite systems, by
the statistics of intergalactic absorption in the spectra of high redshift
objects, and by gravitational lensing of distant galaxies. The first requires
large galaxy samples with precisely measured velocities; the second requires
high S/N spectra of many faint objects to achieve good statistics with a high
sampling density on the sky; the third needs large spectroscopic samples to
the faintest possible limits in order to calibrate the photo-z error
distribution as a function of true redshift.

Dark energy can be constrained by precise measurements of the cosmic expansion
history and of the linear growth of cosmic structure.  Spectroscopy of large
samples of faint galaxies are mandatory to control for systematics in all
currently proposed schemes to make these measurements.  Large-scale
photometric surveys aiming to measure baryon acoustic oscillations or lensing
statistics again require a precise calibration of photo-z errors as a function
of true redshift. Methods based on cluster evolution need spectroscopy to
measure cluster velocity dispersions and to characterize the evolution of
cluster galaxies, in particular, the prevalence of AGN activity.
Supernova-based methods require a good understanding of the evolution of the
parent galaxy population (metallicities, SFRs, dust contents) to constrain
systematic variations in progenitor properties and supernova environments.

The process which generated structure in the early Universe can be constrained
by measuring the linear power spectrum of density fluctuations. The cosmic
microwave background is the prime tool for such studies, but degeneracies
which arise in analyses of CMB data alone can be broken with sufficiently
precise data on the late-time matter distribution. Again gravitational lensing
estimates of the relevant quantities will need exquisite photo-z error
characterizations, while direct estimates from the galaxy distribution will
require a detailed understanding of bias effects which can only be robustly
investigated through large spectroscopic surveys.

\section {A large near-infrared spectroscopic survey 
for the next decade}

As outlined above, the desideratum is a survey which obtains the
spectra of a few times $10^5$ galaxies from the visible into the
near IR at each of a sufficient number of redshift slices that one
can follow the evolution of all interesting populations.
The SDSS main sample contained 300,000 galaxies brighter than
$L_*$ within a volume of $\sim$0.05 h$^{-3}$ Gpc$^3$. One basically
wishes to produce a set of samples comparable in size to this 
from redshift about 2 to the present, covering the epoch during which
most of the baryonic matter was assembled and most of the stars in the 
universe were formed. 

At redshift 2, the effective wavelength of the R band is at about 1.8
microns, the red edge of the H band; working longward of this is
possible but technically very much more difficult, and the survey we
envision covers the range 1-1.8 microns with infrared technology and
0.4-1 micron with optical technology.  As stressed above, one really
needs both in order to study both stellar mass from the observed IR,
rest-frame optical, and star-formation rate from the observed optical,
rest-frame near UV, in the same objects, and we cannot really understand
the properties of galaxies without both. We stress here the infrared
coverage; the optical is much easier and has already received a great
deal of attention.
A reasonable set of redshift shells to focus our thinking might be something 
like:

\vspace {0.2cm}

\begin {tabular} {lccccc}

Boundaries &   0.6-0.8 & 0.8-1.05 &  1.05-1.35 &   1.35-1.65 & 1.65-1.95 \\
Mean z     &      0.7   &   0.9   &  1.2   &  1.5     & 1.8\\     
Area/.05Gpc$^3$ &   140  &    100  &    70  &    60  &     60\\
$\lambda_r$     &   1.07 &    1.19 &   1.38 &   1.57 &     1.76\\
AB(r,L$_*$)       &   20.8 &    21.3 &   22.0 &   22.4 &     22.9\\

\end {tabular}

\vspace {0.2cm}

The AB row is the AB magnitude at rest-frame $r$ (6300\AA) for a simple
evolutionary model corresponding to a bluish $L_*$ galaxy today, and
$\lambda_r$ is the corresponding observed wavelength in microns. The brightness
of the target galaxy obviously depends on its star formation history, but this
is  a pessimistic estimate for most relevant objects.


There are about 300,000 galaxies brighter than an evolution-corrected $L_*$ in
each cell, 1.5 million in all. Whether one would really like to
restrict the sample to intrinsically bright galaxies (for example by color
selection) is
debatable; the sample would be roughly twice as large if one 
simply went to a photometric limit in the smallest region corresponding to the
$L_*$ cut at redshift 2. This would sample galaxies up to 2 magnitudes below
$L_*$ at lower redshifts over a somewhat smaller volume, which 
would certainly be
extremely valuable.

What instrumentation is required to do this? It certainly does not exist at
the present time, but is, we believe, technically feasible (though certainly
not easy.) While
expensive, it is very much cheaper than yet another large general-purpose
telescope, and it would scientifically be of enormously more importance.

The surface density of targets is very high.  Of order a million
galaxies in 100 square degrees is $10^4$ galaxies per square degree.  We
will see that for a plausibly realizable setup, reasonable spectra should
be obtainable in of order four hours. Thus for 3 million galaxies,
a 1000-object spectrograph or array would require of order 3000
pointings of 4 hours each, 12,000 hours.  Accounting for reasonable
weather and avoidance of very bright moon, this is about 2200 nights,
about six years. Clearly there are economic trades between number of
fibers and survey duration, and they need to be carefully made.

The most severe problem with infrared spectroscopy in this wavelength
region is the very bright sky background, which, however, is confined to
a large number of very bright OH airglow lines.  The lines are quite
closely spaced, so one needs to go to quite high spectral resolution to
make use of the dark wavelength regions between the lines to reach faint
distant objects.  The OH spectrum and intensities are very well known;
how dark the sky is between the lines is not so well known, and appears
to be completely dominated by scattering arising from fundamental
grating physics, with a small zodiacal contribution. With reasonable
assumptions about the quality of VPH gratings, the calculated
background brightness is of order 19.5AB/arcsec$^2$ except at the longest
wavelengths, where it rises to about a magnitude brighter because of the
extremely strong lines there.
One must
work at a resolving power in excess of about 4000 in order that some
reasonable fraction (at least $\sim$ 60\%) of the spectrum is reasonably
uncontaminated by the OH emission. 

The technological problems are not the same over the whole wavelength region
1-1.8 $\mu$m , and it might well be advantageous to split it in two, 1-1.34
$\mu$m and 1.34-1.8 $\mu$m.  In the short segment, ordinary optical materials
can be used, and it is not necessary to cool the spectrograph; in the long
region, one can use silicon to enormous advantage as a refracting material,
and it IS necessary to cool the spectrograph. In both, doped HgCdTe detectors
have to be used. In the optical and shortwave IR regions spectrographs very
similar to the SDSS ones or those proposed by Hopkins for WFMOS could be
built. The longwave spectrographs are more problematical, but the UVa
group (Skrutskie and Wilson) are building a high-resolution
H-band instrument, which is a
reasonable (but larger and more complex) model for this instrument.
It is probably best economically and strategically to use a separate
fiber set for each of the three spectrographs rather than split the light,
given the very high target density; multiple visits to each field are necessary in any case.

If we suppose we have available for the survey a 6.5-meter telescope with a
reasonably wide field with a cassegrain F/5 focus (such as the MMT), things
scale quite well to the SDSS spectrographs. We have thought mostly about
fiber systems, which make dealing with the large multiplex capability
possible, though we will need object-sky pairs for chopping.
150 micron
fibers (SDSS used 180) 
translate to about 1.0 arcseconds, probably nearly ideal for the
redshift range in question (this is about 6 kpc over most of the range,
similar to the footprint of the fibers for the SDSS main sample).  We need
about 3000 pixels in the wavelength direction, and could use one and a half
2K$^2$ arrays or contract for a somewhat larger chip.  If we use 300
fibers, the interfiber spacing is equal to the fiber footprint, so each
spectrograph will accommodate 150 object/sky pairs, and we need 7
spectrographs. Sensitivity calculations
indicate that one can obtain a S/N of about 7 per resolution element with
this setup at AB=23 in four hours through most of the IR band, falling to
about 3 at the reddest end. 
Rebinning to R$\sim$1000 and taking the missing data due to the OH into
account gives S/N $\sim$ 11 over most of the range degrading to 5 at the 
reddest end.
With
the noise performance now being reached with IR detectors, one exceeds the
read noise limit in short enough times that the chopping will be efficient.


How much would this cost? A reasonable guess scaled from the APOGEE camera for
the longwave IR spectrograph is about \$1.5M, the shortwave one a
bit cheaper, say \$1M.  The visible could be covered by a WFMOS
survey if that is to happen, but if covered here, about \$2M (double
spectrograph to cover the whole range). If we adopt \$3.5M per spectrograph, we
calculate a total cost of about \$30M for 1000-object capability on an
existing 6.5-m telescope. Needless to say, this survey's worth is severely
compromised unless there are corresponding deep optical and IR 
imaging surveys, deep enough to
obtain multicolor structural parameters for, and choose, 
the spectroscopic targets...
but the area is small enough that this can be accomplished with
existing equipment.

The closest competitor to the instrument proposed is FMOS on Subaru,
with 400 fibers (200 pairs), which is currently being commissioned. FMOS
has a fifth the number of objects on a slightly larger telescope which
is a public facility; clearly a survey of the scope needed is impossible
using it.  

This is clearly {\it not} a detailed instrument proposal. It was intended
to motivate further study to produce a real instrument without which
galaxy evolution studies in the coming decade or two will, we strongly
feel, be severely impaired.

\end {document}